# A field approach for pedestrian movement modelling

Amir Ghorbani[1]

[1] Melbourne School of Engineering, The University of Melbourne
Victoria 3010, Melbourne, Australia
ghorbania@student.unimelb.edu.au

## Abstract

There are different physics-based approaches for analysing pedestrian movement. Physics-based methods like statistical mechanics-based models apply the laws of physics to drive equations for analysing crowd behaviour. This paper will introduce a physics-based approach based on field theory as a new tool for crowd analysis to determine governing differential equations. Formulating the pedestrian movement with differential equations has a primary advantage for data assimilation techniques because some of these methods only work with models with analytical transition functions, which are obtained by incorporating a field approach. Furthermore, the field approach provides more generality since the field could be any scalar field. Several Lagrangians are presented in this work, and the primary purpose was to lay the groundwork for this new type of thinking. Furthermore, as pedestrian movement is mainly unregulated, the approach presented in the paper can be valuable for future development since the Lagrangian could be explicitly obtained for pedestrian movement spaces such as train stations and shopping malls. Finally, we discuss a general approach for predicting the action and how neural networks might play a role, which brings more flexibility and extendibility to our approach.

## 1. Introduction

Pedestrians are an essential part of the transport systems, and simulating their behaviour is crucial for transport facilities management( Nassir et al. 2015; Nassir, Hickman, & Ma 2017). While several modelling approaches for pedestrian behaviour, such as data-driven-based, vision-based, forced-based, and velocity-based, looking at pedestrian movement in terms of the differential equation has a primary advantage for online pedestrian simulation using the data assimilation method. This is because some data assimilation techniques like Extended Kalman Filter(EKF) favour an analytical form of the transition function, which needs to be added in many modelling techniques, especially agent-based ones. (Clay et al. 2020, 2021; Malleson et al. 2020; Ternes et al. 2021). Online simulation can bring numerous advantages, and the interested reader is referred to ( Swarup & Mortveit 2020 ) for more information. Moreover, the formulation of crowd flow in Lagrangian mechanics can be a new tool for crowd motion analysis. In this paper, we are proposing a Lagrangian field approach to analyse crowd behaviour. Certain principles of physics are involved in constructing a Lagrangian; however, there are many choices for the Lagrangian of a system. It is up to experiments(Dias & Sarvi 2016; Shiwakoti, Shi, Ye, & Wang 2015; Shiwakoti, Tay, & Stasinopoulos 2017; Shiwakoti, Tay, Stasinopoulos, & Woolley 2016)to determine whether a Lagrangian is suitable for the system or not. This is especially true for pedestrian systems because they are largely unregulated(Haghani & Sarvi 2018, 2019; Haghani, Sarvi, & Shahhoseini 2019; Haghani, Sarvi, Shahhoseini, & Boltes 2019). While there have been efforts to apply the Lagrangian approach to pedestrian movement (Mukherjee et al. 2015), the main contribution of this paper is starting to construct Lagrangians based on field theory from the most straightforward cases. However, calibration against real-world data is a subject for future research. The proposed methods, especially the one discussed in Sec. [5], have the potential to be calibrated against real-world data, thus being a suitable line of research for future studies.

## 2. Formulation of The Lagrangian

| Notation | Definition |
|---|---|
| $\phi$ | Field |
| L | Lagrangian |
| A | Action |
| V | Potential energy |
| T | Kinetic energy |
| $v$ | Velocity |
| $\rho$ | Density |
| $J_x$ | Flux flow rate ( x direction) |
| $J_y$ | Flux flow rate ( y direction) |

**Table 1 : Notation**

In Lagrangian mechanics, we write Lagrangian (L) for a particle, L=T-V, where T and V are the kinetic and potential energy of the system. Thus, $L := \frac{1}{2}mv^2 - V(x)$ for a single-particle. We apply the same approach for the field and drive a general Lagrangian as below (Susskind Leonard ; Friedman Art 2017) :

$$L := \beta\left[\left(\frac{\partial\phi}{\partial t}\right)^2 - \left(\frac{\partial\phi}{\partial x}\right)^2 - \left(\frac{\partial\phi}{\partial y}\right)^2\right] - V(\phi) \qquad (1)$$

$\beta$ is a constant, and $V(\phi)$ is the potential energy of the field. If we put $\phi = x$, $\beta = 0.5$ and omit the $\left(\frac{\partial\phi}{\partial x}\right)^2, \left(\frac{\partial\phi}{\partial y}\right)^2$ terms; we get the same single-particle Lagrangian. $\beta\left[\left(\frac{\partial\phi}{\partial t}\right)^2 - \left(\frac{\partial\phi}{\partial x}\right)^2 - \left(\frac{\partial\phi}{\partial y}\right)^2\right]$ is the term related to the kinetic energy of the field. Therefore, the action is:

$$A = \int L\, dxdydt \qquad (2)$$

We aim to derive a general scalar field ($\phi$) equation for crowd motion in spacetime. We propose an equation with a general scalar field that can be substituted with other scalar properties like density in the crowd flow. In the crowd flow, no actual forces are acting on the pedestrians (we are not considering the effect of pushing); therefore, in constructing our Lagrangian, we can assume that the potential energy is zero ($V(\phi) = 0$). However, we should define potential functions to include more pedestrian flow characteristics.

Applying the Euler-Lagrangian equations(Susskind Leonard ; Friedman Art 2017; Wald 1984), we get:

$$\sum_\mu \left(\frac{d}{dx^\mu} \frac{\partial L}{\partial \frac{\partial \phi}{\partial x^\mu}}\right) - \frac{\partial L}{\partial \phi} = 0 \,, \mu \in \{0,1,2\} \,\&\, dx^0 = dt, dx^1 = dx, dx^2 = dy \quad (3)$$

$$\Rightarrow 2\beta \left(\frac{\partial^2 \phi}{\partial^2 t} - \frac{\partial^2 \phi}{\partial^2 x} - \frac{\partial^2 \phi}{\partial^2 y}\right) + \frac{\partial V(\phi)}{\partial \phi} = 0 \quad (4)$$

If we assume that potential energy is zero( $V(\phi) = 0$ ), we have:

$$\frac{\partial^2 \phi}{\partial^2 t} - \frac{\partial^2 \phi}{\partial^2 x} - \frac{\partial^2 \phi}{\partial^2 y} = 0 \quad (5)$$

The conservation equation is(Kachroo, Al-Nasur, Wadoo, & Shende 2008):

$$\rho_t + \frac{\partial J_x}{\partial x} + \frac{\partial J_y}{\partial y} = \rho_t + (\rho v)_x + (\rho u)_y = 0 \quad (6)$$

$J_x, J_y$ are flux flow rates in x, y directions. *v* (*x, y, t*) and *u* (*x, y, t*) are the *x*-axis and *y*-axis components of the velocity.

We couple the Eq. (4) with the conservation equation (Eq. (6)) to get a system of PDE equations that can be solved numerically:

$$\begin{cases} \rho_t + (\rho v)_x + (\rho u)_y = 0 \\ 2\beta \left(\frac{\partial^2 \phi}{\partial^2 t} - \frac{\partial^2 \phi}{\partial^2 x} - \frac{\partial^2 \phi}{\partial^2 y}\right) + \frac{\partial V(\phi)}{\partial \phi} = 0 \end{cases} \quad (7)$$

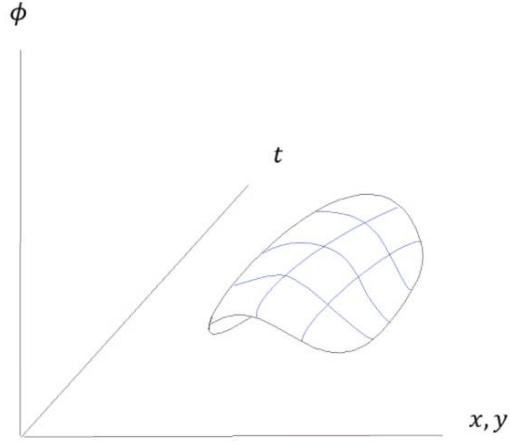

**Figure1: General representation of field over spacetime. x,y are combined spatial axis**

## 3. Potential Energy Function

Drawing an analogy from the social force model(Helbing & Molnár 1995), we define a potential energy function:

$$V \coloneqq V_{repulsive,interaction} + V_{attractive,interaction} + V_{repulsive,obstacle}$$

$$V_{repulsive,interaction} \coloneqq \int_{y \in A} \int_{x \in A} \int_{r'=-\frac{R_0}{2}}^{r'=\frac{R_0}{2}} \int_{r=-\frac{R_0}{2}}^{r=\frac{R_0}{2}} \rho(x-r, y-r', t) V_o e^{-\frac{\sqrt{r^2+r'^2}}{R}} dr\, dr'\, dx\, dy$$

$$V_{attractive,interaction} \coloneqq 0$$

$$V_{repulsive,obstacle} \coloneqq \int_{y \in A} \int_{x \in A} \int_{r'=-\frac{R'_0}{2}}^{r'=\frac{R'_0}{2}} \int_{r=-\frac{R'_0}{2}}^{r=\frac{R'_0}{2}} \rho'(x-r, y-r', t) U_o e^{-\frac{\sqrt{r^2+r'^2}}{R'}} dr\, dr'\, dx\, dy \; ,$$

where $\rho'(x-r, y-r', t) = \rho(x-r, y-r', t)$ if $(x-r, y-r', t) \in$ obstacle area, otherwise $\rho'(x-r, y-r', t) = 0$

The parameters $R_0$, $\mathbb{R}'_0$ are the effective radius of influence for repulsive forces. $V_o$, R, $U_o$ and $\mathbb{R}'$ are defined as (Helbing & Molnár 1995) and $A$ is the area occupied by pedestrians. For simplicity (like (Helbing & Molnár 1995)), we have assumed that there are no attractive forces. Obstacle areas have higher densities ($\rho_{obstacle}$). Therefore:

$$V := \int_{y \in A} \int_{x \in A} \int_{r'=-\frac{R_0}{2}}^{r'=\frac{R_0}{2}} \int_{r=-\frac{R_0}{2}}^{r=\frac{R_0}{2}} \rho(x-r, y-r', t) V_o e^{-\frac{\sqrt{r^2+r'^2}}{R}} dr\, dr'\, dx\, dy\, +$$

$$\int_{y \in A} \int_{x \in A} \int_{r'=-\frac{R_0'}{2}}^{r'=\frac{R_0'}{2}} \int_{r=-\frac{R_0'}{2}}^{r=\frac{R_0'}{2}} \rho'(x-r, y-r', t) U_o e^{-\frac{\sqrt{r^2+r'^2}}{R'}} dr\, dr'\, dx\, dy \quad (8)$$

We calculate the potential energy term in the above integral with this procedure:

1. For a given time t, we go to position (x, y) in the area (A) occupied by pedestrians.
2. we calculate the potential energy ($V_{repulsive,interaction}$ and $V_{attractive,interaction}$) caused by other pedestrians in the neighborhood. ( $x - \frac{R_0}{2} < x < x + \frac{R_0}{2}$, $y - \frac{R_0}{2} < y < y + \frac{R_0}{2}$ )
3. we calculate the potential energy $V_{repulsive,obstacle}$ caused by obstacles in the neighborhood. ( $x - \frac{R_0'}{2} < x < x + \frac{R_0'}{2}$, $y - \frac{R_0'}{2} < y < y + \frac{R_0'}{2}$ )
4. We sum over all (x, y) in the area

## 3. Time as a Field

Let's define the Lagrangian as below ($\phi = t$):

$$L := \beta \left[ \left(\frac{\partial t}{\partial t}\right)^2 - \left(\frac{\partial t}{\partial x}\right)^2 - \left(\frac{\partial t}{\partial y}\right)^2 \right] - V(t) \quad (9)$$

Since adding a constant is allowed in the Lagrangian, we can write the Lagrangian:

$$L := \beta' \left[ \left(\frac{\partial t}{\partial x}\right)^2 + \left(\frac{\partial t}{\partial y}\right)^2 \right] - V(t) \quad (10)$$

Applying the Euler Lagrangian equation, we get:

$$2\beta' \left( \frac{\partial^2 t}{\partial^2 x} + \frac{\partial^2 t}{\partial^2 y} \right) + \frac{\partial V(t)}{\partial t} = 0 \quad (11)$$

Coupling the Eq. (11) with the conservation equation (Eq. (6)) we will have a set of PDEs (Eq. (7)) to track the evolution of crowd motion.

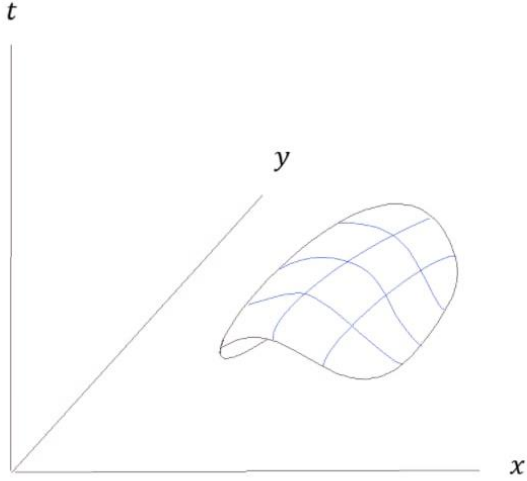

**Figure 2: Time as field**

## 4. Density Field

If we assume $V(\rho) = 0$, we have:

$$\frac{\partial^2 \rho}{\partial^2 t} - \frac{\partial^2 \rho}{\partial^2 x} - \frac{\partial^2 \rho}{\partial^2 y} = 0 \qquad (12)$$

Solving Eq. (12), we get:

$\rho(x,y,t) = \frac{1}{6}\big(6\,c_1 + 6\,y\,c_2 + 6\,t^2\,c_3 + 6\,y^2\,c_3 + 18\,t^2\,y\,c_4 + 6\,y^3\,c_4 + 6\,x^4\,c_5 - 36\,x^2\,y^2\,c_5 + 6\,y^4\,c_5 + 6\,x\,c_6 + 6\,x\,y\,c_7 + 6\,t^2\,x\,c_8 + 6\,x\,y^2\,c_8 + 18\,t^2\,x\,y\,c_9 + 6\,x\,y^3\,c_9 + 6\,t^2\,c_{11} + 6\,x^2\,c_{11} + 6\,t^2\,y\,c_{12} + 6\,x^2\,y\,c_{12} + 18\,t^2\,x\,c_{16} + 6\,x^3\,c_{16} + 18\,t^2\,x\,y\,c_{17} + 6\,x^3\,y\,c_{17} + 6\,t\,c_{26} + 6\,t\,y\,c_{27} + 2\,t^3\,c_{28} + 6\,t\,y^2\,c_{28} + 6\,t^3\,y\,c_{29} + 6\,t\,y^3\,c_{29} + 6\,t\,x^4\,c_{30} - 36\,t\,x^2\,y^2\,c_{30} + 6\,t\,y^4\,c_{30} + 6\,t\,x\,c_{31} + 6\,t\,x\,y\,c_{32} + 2\,t^3\,x\,c_{33} + 6\,t\,x\,y^2\,c_{33} + 6\,t^3\,x\,y\,c_{34} + 6\,t\,x\,y^3\,c_{34} + 2\,t^3\,c_{36} + 6\,t\,x^2\,c_{36} + 2\,t^3\,y\,c_{37} + 6\,t\,x^2\,y\,c_{37} + 6\,t^3\,x\,c_{41} + 6\,t\,x^3\,c_{41} + 6\,t^3\,x\,y\,c_{42} + 6\,t\,x^3\,y\,c_{42} + t^4\,c_{53} - x^4\,c_{53} + 6\,t^2\,y^2\,c_{53} + 6\,x^2\,y^2\,c_{53} + 3\,t^4\,y\,c_{54} - 3\,x^4\,y\,c_{54} + 6\,t^2\,y^3\,c_{54} + 6\,x^2\,y^3\,c_{54} + t^4\,x\,c_{58} + 6\,t^2\,x\,y^2\,c_{58} + x\,y^4\,c_{58} + 6\,t^4\,x\,y\,c_{59} + 6\,t^2\,x^3\,y\,c_{59} + 6\,t^2\,x\,y^3\,c_{59} + 2\,x^3\,y^3\,c_{59} + t^4\,c_{61} + 6\,t^2\,x^2\,c_{61} + x^4\,c_{61} + t^4\,y\,c_{62} + 6\,t^2\,x^2\,y\,c_{62} + x^4\,y\,c_{62} + 3\,t^4\,x\,c_{66} + 6\,t^2\,x^3\,c_{66} + 6\,x^3\,y^2\,c_{66} - 3\,x\,y^4\,c_{66} - 6\,t^3\,x^2\,c_{78} - 6\,t\,x^4\,c_{78} + 6\,t^3\,y^2\,c_{78} + 18\,t\,x^2\,y^2\,c_{78} - 18\,t^3\,x^2\,y\,c_{79} - 18\,t\,x^4\,y\,c_{79} + 6\,t^3\,y^3\,c_{79} + 18\,t\,x^2\,y^3\,c_{79} - 2\,t^3\,x^3\,c_{83} + 6\,t^3\,x\,y^2\,c_{83} - 6\,t\,x^3\,y^2\,c_{83} + 6\,t\,x\,y^4\,c_{83} - 6\,t^4\,x^2\,c_{103} - 6\,t^2\,x^4\,c_{103} + 6\,t^4\,y^2\,c_{103} - 6\,x^4\,y^2\,c_{103} + 6\,t^2\,y^4\,c_{103} + 6\,x^2\,y^4\,c_{103}\big)$

The density function obtained above will be substituted in Eq. (6), and the case is finished. Now, we try to add the potential energy terms and write Eq. (4) for this case.

The new Lagrangian is:

$$L := \beta\left[\left(\frac{\partial\rho}{\partial t}\right)^2 - \left(\frac{\partial\rho}{\partial x}\right)^2 - \left(\frac{\partial\rho}{\partial y}\right)^2\right] - V(\rho) \tag{13}$$

Applying the Euler-Lagrangian equations, we get:

$$2\beta\left(\frac{\partial^2\rho}{\partial^2 t} - \frac{\partial^2\rho}{\partial^2 x} - \frac{\partial^2\rho}{\partial^2 y}\right) + \frac{\partial V(\rho)}{\partial \rho} = 0 \tag{14}$$

Where from Eq. (8) we can write:

$$\frac{\partial V(\rho)}{\partial \rho} = \int_{r'=-\frac{R_0}{2}}^{r'=\frac{R_0}{2}} \int_{r=-\frac{R_0}{2}}^{r=\frac{R_0}{2}} [\frac{\rho_x(x-r,y,t)dx+\rho_y(x,y-r',t)dy+\rho_t(x,y,t)dt-\rho_x(x-r,y,t)dr-\rho_y(x,y-r',t)dr'}{\rho_x(x,y,t)dx+\rho_y(x,y,t)dy+\rho_t(x,y,t)dt} \, e^{-\frac{\sqrt{r^2+r'^2}}{R}} +$$

$$\frac{-1}{\frac{\sqrt{r^2+r'^2}}{R}} e^{-\frac{\sqrt{r^2+r'^2}}{R}}(rdr+r'dr')\rho(x-r,y-r',t)\frac{1}{\rho_x(x,y,t)dx+\rho_y(x,y,t)dy+\rho_t(x,y,t)dt}]V_o dr\, dr' +$$

$$\int_{r'=-\frac{\mathbb{R}'_0}{2}}^{r'=\frac{\mathbb{R}'_0}{2}} \int_{r=-\frac{\mathbb{R}'_0}{2}}^{r=\frac{\mathbb{R}'_0}{2}} [\frac{\rho'_x(x-r,y,t)dx+\rho'_y(x,y-r',t)dy+\rho'_t(x,y,t)dt-\rho'_x(x-r,y,t)dr-\rho'_y(x,y-r',t)dr'}{\rho_x(x,y,t)dx+\rho_y(x,y,t)dy+\rho_t(x,y,t)dt} \, e^{-\frac{\sqrt{r^2+r'^2}}{R}} +$$

$$\frac{-1}{\frac{\sqrt{r^2+r'^2}}{R}} e^{-\frac{\sqrt{r^2+r'^2}}{R}}(rdr+r'dr')\rho'(x-r,y-r',t)\frac{1}{\rho_x(x,y,t)dx+\rho_y(x,y,t)dy+\rho_t(x,y,t)dt}]U_o dr\, dr'$$

Now we can write Eq. (7) for this case.

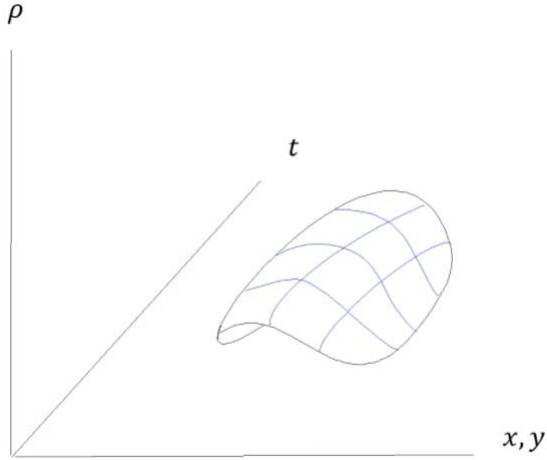

**Figure 3: Density as field**

## 5. Predicting the Lagrangian (Action)

We assume that $\phi(x, y, z)$ has a general form of a polynomial of degree $k$ (with respect to x, y, t):

$$\phi(x, y, t) = \sum_{0 \leq \alpha_0, \alpha_1.\alpha_2 \leq k} \Gamma_{\alpha_0,\alpha_1.\alpha_2} t^{\alpha_0} x^{\alpha_1} y^{\alpha_2} \tag{15}$$

Where $\Gamma_{\alpha_0,\alpha_1.\alpha_2}$ is the coefficient related to $t^{\alpha_0} x^{\alpha_1} y^{\alpha_2}$ term. Therefore :

$$f_\phi(x, y, t) = \beta \left[ \left(\frac{\partial \phi}{\partial t}\right)^2 - \left(\frac{\partial \phi}{\partial x}\right)^2 - \left(\frac{\partial \phi}{\partial y}\right)^2 \right] \tag{16}$$

The kinetic term $f_\phi(x, y, t)$ is polynomial and can be easily integrated. We write Lagrangian and Action as below:

$$L := f_\phi(x, y, t) - V(\phi(x, y, t))$$

$$A = \int L \, dxdydt = \int [f_\phi(x, y, t) - V(\phi(x, y, t))] dxdydt \tag{17}$$

We can fit the data of an experiment to Eq. (15) and determine $\Gamma_{\alpha_0,\alpha_1.\alpha_2}$. Knowing a proper function for $V(\phi(x, y, t)$ like Eq. (8), Then we can calculate the Lagrangian and Action.

For example, if we put $\phi(x, y, t) = \rho(x, y, t)$, We can calculate $V(\rho)$ as a function of x, y, t by using Eq. (8). Therefore: $V(\rho) = g(x, y, t)$ . Now the Action becomes:

$$A(a, b, \Gamma_{\alpha_0,\alpha_1.\alpha_2}) = \int_a^b [f_\rho(x, y, t) - g(x, y, t)] \, dxdydt \tag{18}$$

Where $a$ and $b$ are the start and end boundaries of the action in (x, y, t). We can also calculate the Euler-Lagrangian equations (Eq. (14)) easily and check how well our model works.

We can train a neural network to predict $\Gamma_{\alpha_0,\alpha_1.\alpha_2}$ values based on simulation software. The input variables of the neural network are initial pedestrian positions in space-time x, y, t and the geometry of the area, and the output is $\Gamma_{\alpha_0,\alpha_1.\alpha_2}$. We can go further and write lagrangians for each output of the neural network using Eq. (17) and try to develop a general Lagrangian by comparing and analysing the pattern and form of different Lagrangians.

## 6. Conclusion and future research

In this paper, we presented a field Lagrangian. Pedestrian flows are unregulated and cannot be treated as moving particles, but we have tried to investigate how we can write for pedestrian flows. We can think of other types of Lagrangians with different kinetic and potential energy terms. For example, recalling the Greenshields linear relation for density-velocity, the kinetic energy term can be written as:

$$T := \eta_0 \int \rho V^2 = \eta_0 \int \rho(x,y,t) * (\alpha \rho(x,y,t) + \beta)^2 dx dy \tag{19}$$

Where $\eta_0, \alpha$ and $\beta$ are constants. Therefore, the lagrangian becomes:

$$L := \eta_0 \int \rho(x,y,t) * (\alpha \rho(x,y,t) + \beta)^2 dx dy + V \tag{20}$$

Now we have two sets of Euler-Lagrangian equations for x and y-direction:

$$\begin{aligned}\frac{d}{dt}\frac{\partial L}{\partial \dot{x}} - \frac{\partial L}{\partial x} = 0 \\ \frac{d}{dt}\frac{\partial L}{\partial \dot{y}} - \frac{\partial L}{\partial y} = 0\end{aligned} \tag{21}$$

Substituting the Lagrangian in Eq. (20) to Eq. (21), we get a new set of equations for crowd flow (including the conservation equation). Using these equations (Eq. (19) to Eq. (21)) means that we are looking at pedestrians as particles having kinetic and potential energies, and the specific derivation just mentioned is not a field approach. It is up to experiments to determine how realistic these assumptions are and how we should modify them for pedestrian flow.

.